\documentclass[aip,apl, groupedaddress, preprint]{revtex4-1}

\usepackage{graphicx}

\begin{document}

\title{Suppression of bulk conductivity in InAs/GaSb broken gap composite quantum wells}

\author{Christophe Charpentier}
\email{charpentier@phys.ethz.ch.}
\author{Stefan F\"alt}
\author{Christian Reichl}
\author{Fabrizio Nichele}
\author{Atindra Nath Pal}
\author{Patrick Pietsch}
\author{Thomas Ihn}
\author{Klaus Ensslin}
\author{Werner Wegscheider}
\affiliation{ 
Laboratory for Solid State Physics, ETH Z\"urich, 8093 Z\"urich, Switzerland
}%

\date{\today}

\begin{abstract}
The two-dimensional topological insulator state in InAs/GaSb quantum wells manifests itself by topologically protected helical edge channel transport relying on an insulating bulk. This work investigates a way of suppressing bulk conductivity by using gallium source materials of different degrees of impurity concentrations. While highest-purity gallium is accompanied by clear conduction through the sample bulk, intentional impurity incorporation lead to a bulk resistance over 1\,M$\Omega$. This resistance was found to be independent of applied magnetic fields. Ultra high electron mobilities for GaAs/AlGaAs structures fabricated in a molecular beam epitaxy system used for the growth of Sb-based samples are reported.
\end{abstract}

\maketitle
\begin{figure}
\includegraphics{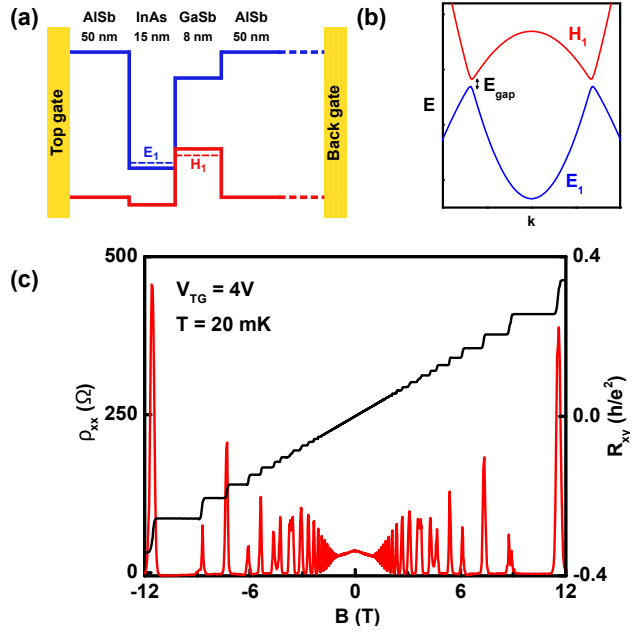}
\caption{(a) Schematic representation of the conduction (blue) and valence (red) band alignments of the CQW region. The lowest (highest) electron (hole) energy levels $E_1$ ($H_1$) are indicated by dashed lines. (b) Dispersion relation of $E_1$ and $H_1$ with the hybridization gap opening at the crossing points of the respective dispersion curves. (c) Magnetotransport data of an InAs/GaSb/AlSb CQW in the electron regime (at a top gate voltage of 4\,V, i.e. an electron density of $2.5~\times~10^{12}$\,cm$^{-2}$). Red: longitudinal resistance, black: transverse resistance.}
\label{fig:fig1}
\end{figure}

Topological insulators (TIs)  are a novel state of matter characterized by topologically protected states forming at the surface of an otherwise insulating system\cite{Kane:2005gb}. In two-dimensional (2D) TIs, this corresponds to helical edge-channels at the edge of an insulating 2D-system, giving rise to the so called quantum spin hall phase. Shortly after the first theoretical prediction of such a system \cite{Bernevig:2006ij}, it was experimentally  realized in HgTe/CdTe quantum wells (QWs)\cite{Konig:2007hs}. The quest for material systems with larger energy gaps protecting the TI state and electric field tunability from metal to a conventional and topological insulator recently lay the focus on InAs/GaSb/AlSb composite quantum wells (CQWs) with their unique broken gap band alignment\cite{Liu:2008zz}, shown in Figure~\ref{fig:fig1}(a). These structures show strong ambipolar behavior, their conductivity can be tuned from electron-like to hole-like by applying an external electric field. From band structure calculations\cite{Liu:2008zz}, a hybridization gap at the crossing of the electron bands from InAs and the hole bands from GaSb is expected at finite densities for adequate layer thickness ratios, see Figure~\ref{fig:fig1}(b). If the Fermi energy is tuned to be inside this gap, transport should be dominated by helical edge channels. First experimental signs of such states in InAs/GaSb CQW samples smaller than the phase coherent length were recently reported\cite{Knez:2010bi, Knez:2011jx, Suzuki:2013ct}. In all studies, the principal issue is the residual conductivity from the bulk which is attributed to sample disorder and obscures the visibility of the dissipationless edge channel transport. If the disorder in the sample cannot be reduced sufficiently to observe pure edge channel transport, an alternative approach would be to adjust the disorder in the sample to such a degree that the mobility of the charge carriers in the bulk is so low that they no longer provide scattering channels between the edge states. Schemes consisting of Si doping at the interface between InAs and GaSb\cite{Du:2013va} or of p-doping of the Al(Ga)Sb barriers surrounding the CQWs\cite{Suzuki:2013ct} have been employed. However, these approaches provide strong scatterers and are thus also prone to strongly influence edge channel transport. This work studies a further strategy of obtaining an insulating sample bulk material by creating a smooth disorder potential using gallium source material with an adequate impurity concentration and thus confirming that the TI phase can be made visible by sufficiently reducing the charge carrier mobility in the bulk.\\ \\
All samples used for this study were grown in the same custom made molecular beam epitaxy (MBE) system equipped with two different gallium effusion cells. The two cells were filled with materials from two different manufacturers, both nominally of highest purity MBE grade. As only metallic impurities are measured to indicate the nominal purity of the material, the quality of the source materials can still vary widely due to non-metallic impurities (as e.g. by their carbon, nitrogen or oxygen content).
The InAs/GaSb/AlSb CQW samples were grown on n-doped $2''$ GaAs substrates insulated from the active region by a 200\,nm low temperature GaAs layer grown at 300\,$^\circ$C. The more than 7\,\% lattice mismatch was accommodated by a 3.2\,nm wide AlSb nucleation layer\cite{Tuttle:1989ti} followed by a 1.5\,$\mu$m GaSb/AlGaSb buffer layer similar to those studied by Nguyen et al.\cite{Nguyen:1993tj}, both grown at 540\,$^\circ$C. The active region consists of the 8\,nm GaSb and 15\,nm InAs QW sandwiched between two 50\,nm AlSb barriers. Before the growth of the InAs layer, the temperature was lowered to 450\,$^\circ$C and the shutter sequence proposed by Tuttle et al.\cite{Tuttle:1990vq} was used to produce InSb bonds at the interface between InAs and AlSb/GaSb. For both group V elements, a valved cracker cell was used. The Sb cracker zone was operated at a temperature of 700\,$^\circ$C, giving an approximately 50:50 mixture of Sb$_2$ and Sb$_4$\cite{Rouillard:1995wv} and which prevents cross-contamination by As deposited near the Sb cell. We used uncracked As$_4$ as arsenic source, the cracker zone being heated to 400\,$^\circ$C. All layers were grown under group V overpressure, with beam equivalent pressures (BEPs) of $1.75\times10^{-6}$\,torr for Sb$_2$/Sb$_4$ and $8.5\times10^{-6}$\,torr for As$_4$ respectively. On both samples, identical 50\,$\mu$m $\times$ 25\,$\mu$m Hall bars were etched by conventional photolithography and plasma etching. The devices were then covered by a 200\,nm Si$_3$N$_4$ layer deposited by plasma enhanced chemical vapor deposition followed by a metallic top gate to tune the electron and hole densities.

Electronic transport experiments were performed in a pumped bath cryostat at a temperature of 1.3\,K at magnetic fields up to 7\,T and in a $^3$He/$^4$He dilution refrigerator at 20\,mK at magnetic fields up to 12\,T using standard low-frequency lock-in technique. Photoluminescence spectra were taken at 4.2\,K using a HeNe laser as an excitation source at a wavelength of 633\,nm. \\ \\
\begin{figure}
\includegraphics{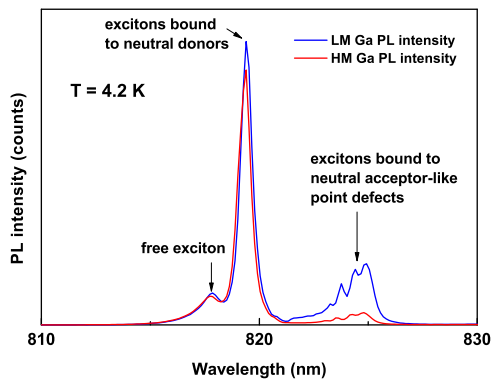}
\caption{Near-band-edge PL spectra of GaAs structures grown using both gallium sources. Blue: low mobility gallium, red: high mobility gallium. The peaks are labeled according to references \onlinecite{Ashen:1975wu, Kunzel:1980co}.}
\label{fig:pl}
\end{figure}
The electron mobility of remotely doped 2D-electron systems is a direct measure for the quality of the source materials as impurities contained in these source materials will be incorporated in the samples and thus lead to additional scattering. The AlGaAs/GaAs double-sided doped  QW structures used for quality assessment are known to show mobilities over $35\times10^6$\,cm$^2$/Vs if grown in a MBE system optimized for the growth of ultra high mobility AlGaAs/GaAs samples\cite{Umansky:2009hb, Pfeiffer:2003gj}. In our system also used for the growth of antimony based samples, employing such a structure, we could reproducibly achieve low-temperature ($T\approx1$\,K) mobilities of $14\times10^6$\,cm$^2$/Vs at a density of $3.1~\times~10^{11}$\,cm$^{-2}$ for one charge of gallium source material, which we will subsequently refer to as high mobility (HM) gallium. In contrast, if using the second gallium source material (referred to below as low mobility (LM) gallium), we did not reach mobilities higher than $1.5\times10^6$\,cm$^2$/Vs at a density of $2.7~\times~10^{11}$\,cm$^{-2}$. This demonstrates that the LM gallium is of inferior quality in terms of impurity concentrations compared to the HM gallium but still allows to grow samples with reasonably high mobilities. This finding is confirmed by comparing the PL spectra shown in Figure~\ref{fig:pl}, where the off band gap peaks linked to impurity assisted absorption are higher by a factor of 8 for the LM gallium compared to the HM gallium. We want to point out that we were able to grow AlGaAs/GaAs samples showing electron mobilities in excess of $10^7$\,cm$^2$/Vs after the growth of antimony based samples. This proves that the presence of Sb in the system does not influence the quality of high-purity arsenide based structures at this level.

Figure~\ref{fig:fig1}(c) shows magneto-transport data of the device fabricated using HM gallium in the electron regime (at a top gate voltage of +4\,V, corresponding to a density of $2.5~\times~10^{12}$\,cm$^{-2}$). The clear hall plateaus and Shubnikov-de Haas oscillations together with a high electron mobility of 300,000\,cm$^2$/Vs at a density of $8~\times~10^{11}$\,cm$^{-2}$ demonstrate the high CQW quality. Samples grown using LM gallium only show a mobility of 8,000\,cm$^2$/Vs at a density of $8.1~\times~10^{11}$\,cm$^{-2}$, i.e. lower by more than an order of magnitude, similar to the AlGaAs/GaAs reference samples. The sharp drop in electron mobility in the electron transport regime only due to higher impurity concentration in the GaSb layer indicates that the electron wave function must extend considerably into the GaSb part of the CQW. For comparison, using LM Ga, electron mobilities as high as 140,000\,cm$^2$/Vs at densities of $9.7~\times~10^{11}$\,cm$^{-2}$ could be achieved in conventional InAs/AlSb QWs where the electron wave function does not reside in a Ga-containing layer.

\begin{figure}
\includegraphics{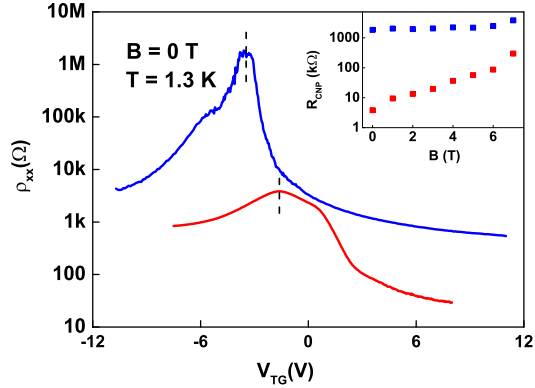}
\caption{Longitudinal resistance of HM Ga (red) and LM Ga (blue) InAs/GaSb/AlSb CQW samples in function of top gate voltage at zero perpendicular magnetic field. The dashed lines indicate the position of the CNP at which $R_{\rm CNP}$ was determined. Inset: $R_{\rm CNP}$ for different B-fields from 0 to 7 T.}\label{fig:peaks}
\end{figure}

We now focus on the effect of the gallium purity on the transport properties of InAs/GaSb CQWs at the charge neutrality point (CNP) where the electron and hole densities are equal and the hybridization gap opens. The sample grown using HM gallium shows a resistance two orders of magnitude higher than the lowest resistance in the high electron density regime at positive top gate voltages. However, the resistivity never rises above 2\,k$\Omega$ even at the CNP. The sample fabricated using the LM gallium shows a far higher increase in resistance, over more than 4 orders of magnitude to 1.5\,M$\Omega$ at the CNP, indicating a truly insulating bulk. In agreement with our previous findings, we observe a qualitatively different transport behaviors for the two samples when a perpendicular magnetic field is applied. As shown in the inset of Figure~\ref{fig:peaks} where we display the longitudinal resistance at the CNP $R_{\rm CNP}$ of both samples for different magnetic fields, for the HM gallium sample, $R_{\rm CNP}$ increases by a factor of 30 when the magnetic field strength is swept from 0 to 7\,T whereas the LM gallium sample shows hardly any dependence of $R_{\rm CNP}$ on the magnetic field. This can be understood as being due to the strong localization at impurities in the LM gallium samples which is much stronger than the localization effect of the B-field.\\ \\
We studied the properties of InAs/GaSb/AlSb CQWs grown using Ga sources of different impurity levels. Even structures fabricated from source materials of highest purity that show very good electron transport properties have a finite density of bulk states in the gap which makes the observation of the helical edge channels and their use for applications as the observation of Majorana fermions\cite{Mi:2013bd} difficult. By using Ga source material with a moderate impurity concentration, we obtained CQW samples with very low bulk conductivity. This will allow to observe edge channel transport in adequately sized Hall bars. In contrast to other schemes tested so far, our method should not significantly alter the edge channel transport properties of the CQW. Another way to implement this strategy would be the controlled addition of isoelectric impurities, e.g. a very small fraction on the order of 1\% of Al to the GaSb CQW layer.

\begin{acknowledgments}
The authors acknowledge financial support by the Swiss National Science Foundation (SNF) and the NCCR QSIT (National Competence Center in Research - Quantum Science and Technology), 
\end{acknowledgments}

\end{document}